\documentstyle[preprint,tighten,aps,12pt,epsf]{revtex}
\draft
\begin{document}
\title{ Radiative corrections to the differential decay rate of polarized
orthopositronium  }
\author{Dmitriy N. Matsukevich\footnote{E-mail: 
mats@inp.belpak.minsk.by} and Oleg N. Metelitsa}
\address{Institute of Nuclear Problems, Belarusian State University \\
St. Bobryiskaya 11, 220050, Minsk, Republic of Belarus }
\date{\today}
\maketitle
\begin{abstract}
The order - $\alpha$ radiative corrections to the differential decay rate
of polarized orthopositronium are obtained. Their influences on the three
photons coincidence rate as a function of positronium polarization is
considered.
\end{abstract}

\narrowtext
Positron experiments for QED testing are of interest at present due to
discrepancy between calculations and the most precise measurements of
orthopositronium decay rate~\cite{Gidly,Gidly2}. Although the recent
experiment~\cite{Asai} shows
good agreement with the order-$\alpha$ theoretical results this problem is
certainly exist and new measurements are necessary. One of the
possible kinds of experiments is the measurements of angular
distribution of the decay photons formed by polarized positronium
proposed in ref.~\cite{mtl}.  In this experiment a three-photon
coincidence rate will be measured as a function of decay plane
orientation. It allows to determine the products of annihilation
amplitudes corresponding to the different positronium spin states.
Usual measurements of positronium decay rate only give information
about the sum of decay amplitudes squared.

In this Letter we have calculated radiative corrections to the differential
decay rate of polarized orthopositronium. Until the present time the similar
calculations were carried out for nonpolarized positronium
only~\cite{str_holt,cls,cls2,big_adk,adk2,Burichen,adk_last}. The expression
for differential decay rate of orthopositronium is
\begin{eqnarray}
d\Gamma &=&
\frac{1}{4m_{e}} \frac{d^{3}k_{1}}{2\omega_{1}}
\frac{d^{3}k_{2}}{2\omega_{2}} \frac{d^{3}k_{3}}{2\omega_{3}}
\frac{\delta(P-k_{1}-k_{2}-k_{3})}{(2\pi)^{4}} \nonumber\\
& & \times
\sum_{\epsilon_{1},\epsilon_{2},\epsilon_{3}}
\frac{1}{3!}
\sum_{m,n}
T(m) \rho_{m,n} T^{*}(n) .
\end{eqnarray}
Here $m_{e}$ is the electron mass, $P=(0,2m_{e})$ is the positronium
rest-frame energy-momentum vector, $k_{i}$ is energy momentum vector
of final state photon with energy $\omega_{i}$ and polarization
vector $\epsilon_{i}$ , $\rho_{m,n}$ is the positronium density
matrix and $T(m)$ is the decay amplitude in $m$ spin state.

When $\delta$ function was integrated out expression (1) took the form
\begin{equation}
d \Gamma=\sum_{m,n}
\rho_{m,n} F_{n,m} d \omega_{1} d \omega_{2}
\frac{do_{1} d \phi_{12}}
{8 \pi^2}  ,
\end{equation}
where solid angle $do_{1}$ contains the first photon
momentum direction, $\phi_{12}$ is the angle of decay plane rotation
around $\vec k_{1}$,  and $F_{n,m}$ is proportional to the positronium decay
matrix~\cite{mtl}
\begin{equation}
F_{n,m} =\frac{1}{2^{7} \pi^{3} m_{e}}
\sum_{\epsilon_{1},\epsilon_{2},\epsilon_{3}}
\frac{1}{3!} T^{*}(n) T(m).
\end{equation}
Polarized positronium decay is described by five parameters. Three angles
define orientation of decay plane and energies of two photons provide
information about relative orientation of photon momenta in this plane.
Elements of decay matrix obviously depend on the choice of reference
frame.  In our calculation we fix the reference frame as follows.
The $z$ axis is directed perpendicularly to the decay plane. The axis
$x$ goes along the momentum of the first photon $\vec k_{1}$ and
projection of $\vec k_{2}$ to the axis $y$ is positive. Projection
$m_{z}$ of positronium spin to the axis $z$ is used to describe the
positronium spin state. When choosing the label of photon momenta we
expect the conditions $\omega_{1}>\omega_{2}>\omega_{3}$ to be met.
The choice of reference frame described above does not affect the
generality of our results because one can always rewrite the
positronium density matrix in our reference frame using angular
moment transformation properties.

Elements of the decay matrix are not independent. Eq.(3) shows that
$F_{ij}=F^{*}_{ji}$. Due to invariance under
CP-transformations~\cite{Deutsch} $F_{1,0}=F_{-1,0}=0$. The CPT
invariance requires~\cite{Deutsch2} that $F_{1,1}=F_{-1,-1}$. Thus,
only two real $F_{1,1}$ and $F_{0,0}$ and one complex number
$F_{1,-1}=|F_{1,-1}|e^{i \varphi}$ are necessary to describe the
polarized positronium decay.

The eigenvalues and eigenvectors of the decay matrix can be
used instead of matrix components. This approach seems more useful
because the eigenvalues and eigenvalues do not depend on the choice of
reference frame. The eigenvalues can be expressed in terms of
decay matrix elements
$ \Lambda_{1}=F_{0,0}$,
$ \Lambda_{2,3}=F_{1,1} \pm |F_{1,-1}| $.
One of the eigenvectors $\vec e_{1}=\vec e_{z}$ is
always perpendicular to the decay plane and two others
\begin{eqnarray}
\vec e_{2}&=&\vec e_{x} sin( \varphi/2) + \vec e_{y} cos( \varphi/2),
\nonumber\\
\vec e_{3}&=&\vec e_{x} cos( \varphi/2) - \vec e_{y} sin( \varphi/2)
\end{eqnarray}
lay in this plane. If one knows the eigenvalues and eigenvectors he
can always represent the decay matrix as follows
\begin{equation}
F_{ij}=
\Lambda_{1}  e_{1i}  e^{*}_{1j} +
\Lambda_{2}  e_{2i}  e^{*}_{2j} +
\Lambda_{3}  e_{3i}  e^{*}_{3j}.
\end{equation}

We have calculated the decay
amplitudes as a function of photon energies and polarizations of photons and
positronium. Then elements of the decay matrix were found as a sum of
corresponding products of amplitudes over photons polarization. Similar
method was used before by Burichenko \cite{Burichen} and Adkins
\cite{adk_last} in their evaluation of order-$\alpha^{2}$ corrections to the
positronium decay rate coming from the squares of one loop amplitudes.

\begin{figure}
\vspace{1cm}
\epsfxsize = 14cm
\centerline{\epsfbox{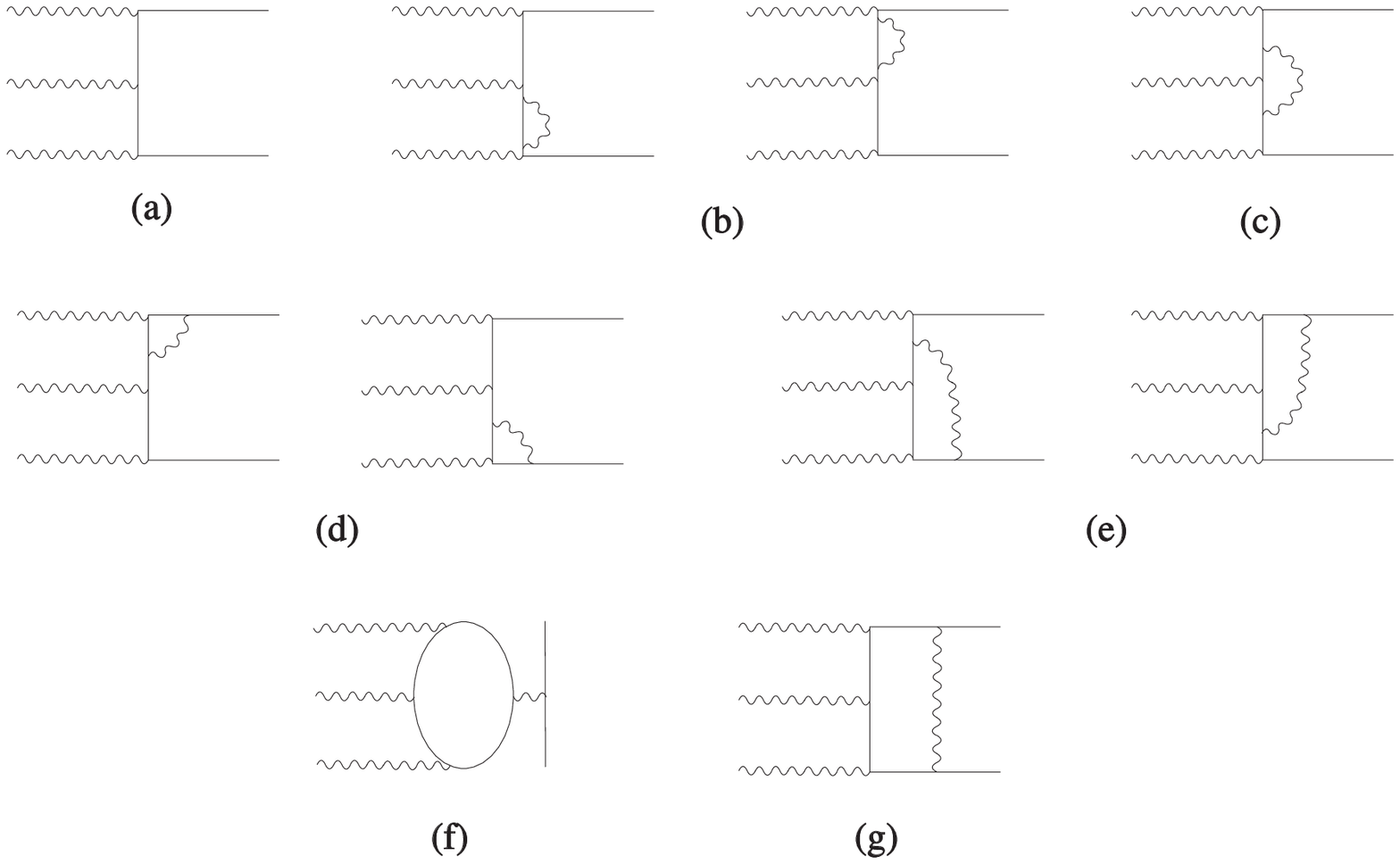}}
\vspace{1cm}
\caption{Feynman graphs contributing to the orthopositronium
decay in the lowest order (a) and in the order-$\alpha$ radiative corrections
(b-j). They are self-energy (b), inner vertex (c), outer vertex (d), double
vertex (e), annihilation (f) and binding (g).}
\label{figa}
\end{figure}

Both Feynman and three-dimensional transverse gauges
of decay photons were used in our calculations.
Virtual photon propagator was taken in Feynman gauge.
The graphs contributing to the lowest order decay amplitudes (a) and
order-$\alpha$ radiative corrections (b-f) are shown on Fig.~\ref{figa}.
Contributions of five diagrams with self-energy and vertex operators inside
were obtained analytically.  Self-energy and outer vertex graphs were
evaluated using well known expressions for one half on mass shell vertex
operator~\cite{axiezer} and self-energy operator~\cite{berest}.  The
analytical expression for off mass shell vertex was obtained to evaluate the
inner vertex graph. Both Feynman parameters technique and direct loop
integration in double vertex graph were used. In the latter method
$k_{0}$ integration was done by poles and remaining three dimensional
integrals were calculated numerically as described in~\cite{cls}. For the
annihilation diagram contribution we followed the treatment proposed
in~\cite{kar_plus,Israel}.  One integration over Feynman parameters was
performed analytically. The remaining two dimensional integrals were computed
numerically. Direct loop integration was carried out to evaluate the binding
graph. Averaging over directions specified by $\vec k$ and $- \vec k$ was
performed for better integral convergence and terms proportional to $\sim
\alpha /v$ were subtracted as described in~\cite{cls,cls2}. Monte Carlo
integration was used for numerical calculations. Gamma matrix products were
evaluated numerically. The program of symbolic manipulations REDUCE was
used for computation.

The eigenvalues of $F_{kl}$ in the lowest order have the form~\cite{mtl}
\begin{eqnarray}
\Lambda^{0}_{2}&=&N(3R-2C)/2, \nonumber\\
\Lambda^{0}_{3}&=&N(-R+2C)/2, \nonumber\\
\Lambda^{0}_{1}&=&\Lambda^{0}_{2}+\Lambda^{0}_{3}=NR.
\end{eqnarray}
Here $N$ is the normalization constant
\begin{equation}
N=3 \Gamma_{0}/[8 m^{2}_{e}(\pi^{2}-9)],
\end{equation}
\begin{equation}
\Gamma_{0} =\frac{2}{9 \pi}(\pi^{2}-9) m \alpha^{6}
\end{equation}
 is the lowest order decay rate,
$$ R=(1-\vec n_{1} \cdot \vec n_{2})^{2}+
(1-\vec n_{1} \cdot \vec n_{3})^{2}+(1-\vec n_{2} \cdot \vec n_{3})^{2} $$
\begin{eqnarray}
C &=& (1-\vec n_{1} \cdot \vec n_{2})(1-\vec n_{1} \cdot \vec n_{3})+
 (1-\vec n_{1} \cdot \vec n_{2}) (1-\vec n_{2} \cdot \vec n_{3}) \nonumber\\
& &  +
(1-\vec n_{1} \cdot \vec n_{3})(1-\vec n_{2} \cdot \vec n_{3}) ,
\end{eqnarray}
$\vec n_{i}=\vec k_{i}/ \omega_{i}$, and their scalar products $ \vec n_{1}
\cdot \vec n_{2}= (\omega^{2}_{3}- \omega^{2}_{1} - \omega^{2}_{2}) / { 2
\omega_{1} \omega_{2} }$, etc.
Angle $\varphi$ in this case can be expressed as follows
\begin{eqnarray}
\tan \varphi^{0} &=& - 2 (n_{3x} n_{3y} (1-\vec n_{1} \cdot \vec n_{2})^{2}
            + n_{2x} n_{2y} (1-\vec n_{1} \cdot \vec n_{3})^{2} \nonumber\\
       & & + n_{1x} n_{1y} (1-\vec n_{2} \cdot \vec n_{3})^{2})/ \nonumber\\
 & & ((n^{2}_{3x}-n^{2}_{3y})(1-\vec n_{1} \cdot \vec n_{2})^{2} \nonumber\\
 & & +(n^{2}_{2x}-n^{2}_{2y})(1-\vec n_{1} \cdot \vec n_{3})^{2} \nonumber\\
 & & +(n^{2}_{1x}-n^{2}_{1y})(1-\vec n_{2} \cdot \vec  n_{3})^{2})
\end{eqnarray}
We assume here that photons momenta lay in the plane $xy$.

We have calculated the changes of eigenvalues and the angle of
eigenvectors rotation around the normal to the decay plane due to
radiative corrections. Our results are presented in Table \ref{firsttab}.
Their errors correspond to $\pm 1$ in the last digit. The radiative
corrections cause significant (up to 2-5 \%) decrease in the decay matrix
eigenvalues in comparison with the lowest-order results and turn the decay
matrix eigenvectors around the decay plane normal. Changes in eigenvalues
due to radiative corrections are almost uniform in the ''center" of phase
space. The anisotropy degree $\eta_{ik}=\Lambda_{i} /
\Lambda_{k}$~\cite{mtl} changes significantly (1-2 \%) near the
''edge" of phase space only. These changes are essentially smaller
than the corrections to the differential decay rate. Radiative
corrections to the differential decay rate averaged over positronium
polarization are in a good agreement with previous
results~\cite{big_adk}.

This work was supported by the Fundamental Research Foundation of the
Republic of Belarus.


\widetext
\begin{table}
\caption{ Order-$\alpha$ radiative corrections to the decay matrix
eigenvalues and change of eigenvectors rotation angle around decay plane
normal due to radiative corrections.}
\label{firsttab}
\begin{tabular}[t]{cccccc|cccccc}
&&\multicolumn{3}{c}
  {$-\frac{\Lambda_{i}-\Lambda^{0}_{i}}{ \Lambda^{0}_{i}} \times 100$}
&&&&
\multicolumn{3}{c}
  {$-\frac{\Lambda_{i}-\Lambda^{0}_{i}}{ \Lambda^{0}_{i}} \times 100$}
&\\
  $ \omega_{1}/m_{e} $ &
  $ \omega_{2}/m_{e} $ &
  $ i=1 $ &
  $ i=2 $ &
  $ i=3 $ &
  ${\frac{\varphi-\varphi^{0}}{2},}^{\circ}$ &
  $ \omega_{1}/m_{e} $ &
  $ \omega_{2}/m_{e} $ &
  $ i=1 $ &
  $ i=2 $ &
  $ i=3 $ &
  ${\frac{\varphi - \varphi^{0}}{2},}^{\circ}$ \\
\tableline
  0.0323& 0.9979&  4.8&  4.8& 4.8&  0.27  & 0.2906& 0.9812& 2.46& 2.45& 2.33& 0.066 \\
  0.0344& 0.9937&  4.5&  4.3& 5.0&  0.54  & 0.3094& 0.9438& 2.40& 2.35& 2.34& 0.145 \\
  0.0365& 0.9896&  4.3&  3.8& 5.3&  0.71  & 0.3281& 0.9063& 2.37& 2.25& 2.39& 0.24  \\
  0.0385& 0.9854&  4.2&  3.4& 5.6&  0.65  & 0.3469& 0.8687& 2.35& 2.13& 2.47& 0.31  \\
  0.0406& 0.9812&  4.1&  3.0& 5.7&  0.25  & 0.3656& 0.8313& 2.32& 2.03& 2.52& 0.17  \\
  0.0969& 0.9937& 3.25& 3.23& 3.2& 0.15  & 0.3552& 0.9771& 2.39& 2.38& 2.24& 0.047 \\
  0.1031& 0.9812& 3.09& 2.98& 3.3& 0.31  & 0.3781& 0.9312& 2.34& 2.31& 2.24& 0.109 \\
  0.1094& 0.9688& 2.98& 2.70& 3.44& 0.44  & 0.4010& 0.8854& 2.31& 2.23& 2.26& 0.20 \\
  0.1156& 0.9563& 2.91& 2.41& 3.62& 0.43  & 0.4240& 0.8396& 2.30& 2.15& 2.31& 0.30  \\
  0.1219& 0.9438& 2.86& 2.22& 3.70& 0.18  & 0.4469& 0.7937& 2.27& 2.06& 2.36& 0.20  \\
  0.1615& 0.9896& 2.80& 2.78& 2.7&  0.11  & 0.4198& 0.9729& 2.35& 2.34& 2.19& 0.029 \\
  0.1719& 0.9688& 2.70& 2.62& 2.77& 0.23  & 0.4469& 0.9187& 2.30& 2.28& 2.17& 0.071 \\
  0.1823& 0.9479& 2.63& 2.42& 2.87& 0.34  & 0.4740& 0.8646& 2.28& 2.23& 2.18& 0.15  \\
  0.1927& 0.9271& 2.58& 2.22& 3.00& 0.35  & 0.5010& 0.8104& 2.27& 2.17& 2.20& 0.30  \\
  0.2031& 0.9063& 2.54& 2.07& 3.07& 0.16  & 0.5281& 0.7563& 2.25& 2.11& 2.26& 0.34  \\
  0.2260& 0.9854& 2.59& 2.58& 2.48& 0.087 & 0.4844& 0.9688& 2.32& 2.32& 2.17& 0.010 \\
  0.2406& 0.9563& 2.52& 2.46& 2.50& 0.18  & 0.5156& 0.9063& 2.28& 2.27& 2.14& 0.025 \\
  0.2552& 0.9271& 2.46& 2.31& 2.58& 0.28  & 0.5469& 0.8438& 2.27& 2.24& 2.13& 0.07 \\
  0.2698& 0.8979& 2.43& 2.14& 2.67& 0.33  & 0.5781& 0.7813& 2.25& 2.21& 2.14& 0.19 \\
  0.2844& 0.8687& 2.41& 2.03& 2.74& 0.16  & 0.6094& 0.7188& 2.24& 2.16& 2.18& 0.7 \\
\end{tabular}
\end{table}


\begin{references}
\bibitem{Gidly}{J.S. Nico, D.W. Gidley, A. Rich and P.W. Zizewitz,
Phys. Rev. Lett. {\bf 65}, 1344 (1990).}
\bibitem{Gidly2}{J.S. Westbrook, D.W. Gidley, R.S. Conti, and A. Rich,
Phys. Rev. A {\bf 40}, 5849 (1989).}
\bibitem{Asai} {S. Asai, S. Orito, and N, Shinohara,
Phys. Lett. B {\bf357}, 475 (1995).}
\bibitem{mtl}{V.G. Baryshevsky and O.N. Metelitsa, Acta Physica Polonica
{\bf 88}, 73 (1995).}
\bibitem{str_holt}{M.A. Stroscio and J.M. Holt, Phys. Rev. A {\bf10}, 749
(1974).}
\bibitem{cls}{W.E. Caswell, G.P. Lepage and J. Sapirstein,
Phys. Rev. Lett. {\bf 38}, 488 (1977).}
\bibitem{cls2}{W.E. Caswell and G.P. Lepage, Phys. Rev. A
{\bf 20}, 36 (1979).}
\bibitem{big_adk} { G.S. Adkins, Annals of Physics {\bf 146}, 78 (1983).}
\bibitem{adk2} {G.S. Adkins, A.A. Salahuddin and K.E. Schalm,
Phys. Rev. A {\bf 45}, 7774 (1992).}
\bibitem{Burichen} {A.P. Burichenko, Yad. Fiz. {\bf 56}, 123 (1993)
[Phys. At. Nucl. {\bf56}, 640 (1993)].}
\bibitem{adk_last} {G.S. Adkins, Phys. Rev. Lett. {\bf 76}, 4903 (1996).}
\bibitem{Deutsch} {W. Bernreuther, U. Low and O. Nachtmann, Hyperfine
Interactions {\bf 44}, 139 (1988).}
\bibitem{Deutsch2} {W. Bernreuther and O. Nachtmann, Z. Phys. C {\bf 11},
235 (1981); B.K. Arbic {\it et al.}, Phys. Rev. A {\bf 37}, 3189 (1988).}
\bibitem{axiezer} {A.I. Akhiezer, V.B. Berestestkii { \it Quantum
Electrodynamics}, Interscience Publishers, New York, London, Sydney,
1965.}
\bibitem{berest} {V.B. Berestetskii, E.M. Lifshiz and L.P. Pitaevskii
{ \it Quantum Electrodynamics}, {\it Landau and Lifshits Course of
Theoretical Physics}, vol. 4, New York, Pergamon, 1982.}
\bibitem{kar_plus} {R.P. Karplus, M. Neuman, Phys. Rev. {\bf 80}, 380 (1950).}
\bibitem{Israel} {Y. Shima, Phys. Rev. {\bf 142}, 944 (1966).}
\end{references}
\end{document}